\documentclass[twocolumn,showpacs,prb,amsmath,amssymb,superscriptaddress]{revtex4}

\usepackage{graphicx}
\usepackage{dcolumn}
\usepackage{bm}

\begin{document}

\preprint{APS/123-QED}

\title{
Anomalous momentum dependence of the multiband electronic structure of
FeSe$_{1-x}$Te$_x$ superconductors induced by atomic disorder
}

\author{Takaaki Sudayama}
\affiliation{Department of Physics, University of Tokyo, 5-1-5 Kashiwanoha, Kashiwa, 
Chiba 277-8561, Japan}
\author{Yuki Wakisaka}
\affiliation{Department of Physics, University of Tokyo, 5-1-5 Kashiwanoha, Kashiwa, 
Chiba 277-8561, Japan}
\author{Daiki Ootsuki}
\affiliation{Department of Physics, University of Tokyo, 5-1-5 Kashiwanoha, Kashiwa, 
Chiba 277-8561, Japan}
\author{Takashi Mizokawa}
\affiliation{Department of Complexity Science 
and Engineering, University of Tokyo, 5-1-5 Kashiwanoha, Kashiwa, 
Chiba 277-8561, Japan}
\affiliation{Department of Physics, University of Tokyo, 5-1-5 Kashiwanoha, Kashiwa, 
Chiba 277-8561, Japan}
\author{Naurang L.~Saini}
\affiliation{Department of Physics, Universit\'a di Roma "La Sapienza",
Piazzale Aldo Moro 2, 00185 Roma, Italy}
\affiliation{Department of Complexity Science 
and Engineering, University of Tokyo, 5-1-5 Kashiwanoha, Kashiwa, 
Chiba 277-8561, Japan}
\author{Masashi Arita}
\affiliation{Hiroshima Synchrotron Radiation Center, Hiroshima University, 
Higashihiroshima, Hiroshima 739-0046, Japan}
\author{Hirofumi Namatame}
\affiliation{Hiroshima Synchrotron Radiation Center, Hiroshima University, 
Higashihiroshima, Hiroshima 739-0046, Japan}
\author{Masaki Taniguchi}
\affiliation{Hiroshima Synchrotron Radiation Center, Hiroshima University, 
Higashihiroshima, Hiroshima 739-0046, Japan}
\affiliation{Graduate School of Science, Hiroshima University, 
Higashihiroshima, Hiroshima 739-8526, Japan}
\author{Takashi Noji}
\affiliation{Department of Applied Physics, Tohoku University, Sendai 980-8579, Japan}
\author{Yoji Koike}
\affiliation{Department of Applied Physics, Tohoku University, Sendai 980-8579, Japan}

\date{\today}

\begin{abstract}
When periodicity of crystal is disturbed by atomic disorder, 
its electronic state becomes inhomogeneous and band dispersion is obscured. 
In case of Fe-based superconductors, disorder of chalcogen/pnictogen height 
causes disorder of Fe 3$d$ level splitting. 
Here, we report an angle-resolved photoemission spectroscopy study
on FeSe$_{1-x}$Te$_x$ with the chalcogen height disorder, 
showing that the disorder affects the Fe 3$d$ band dispersions 
in an orbital-selective way instead of simple obscuring effect.
The reverse of the Fe 3$d$ level splitting due to the chalcogen height 
difference causes the splitting of the hole band with Fe 3$d$ $x^2-y^2$ 
character around the $\Gamma$ point. 
\end{abstract}

\maketitle

\newpage


The inhomogeneous distributions of spin, charge, lattice, and gap magnitude 
discovered in high-Tc cuprates \cite{1,2,3} have inspired tremendous research 
activities on the relationship between the inhomogeneity and 
the high-$T_c$ superconductivity. The inhomogeneous electronic state 
is deeply related to the phase competition or the quantum critical phenomenon 
which plays a central role for the emergence of high-$T_c$ superconductivity. 
On the other hand, the discoveries of high-$T_c$ superconductivity 
in Fe pnictides and chalcogenides have created a new wave of research activities 
on the high-$T_c$ superconductors. \cite{7} 
The Fe-based superconductors are based on a multi-orbital system in which 
several Fermi surfaces with different Fe 3$d$ orbital character are 
responsible for the superconductivity while the cuprate basically has 
single Fermi surface mainly derived by the Cu 3$d$ $x^2-y^2$ orbital. 
Therefore, although the competition between the magnetism and superconductivity 
is commonly seen in the cuprates and the Fe pnictides/chalcogenides, 
the inhomogeneity discovered in the Fe pnictides/chalcogenides could be 
different from that in the cuprates due to the multi-orbital character. 

$\alpha$-FeSe has the simplest crystal structure (anti-PbO type structure)
among the Fe-based superconductors as displayed in Figs. 1(a) and (b)].
While $\alpha$-FeSe shows superconductivity with $T_c$ $\sim$ 8 K, \cite{8}
$\alpha$-FeTe is not superconducting and, instead, shows incommensurate 
spin density wave accompanied by structural transition from tetragonal 
to orthorhombic phase. \cite{9} The spin density wave of FeTe 
exhibits a stripe-type spin modulation, the direction of which is  
different from that of the FeAs-based parent compounds such as LaFeAsO and 
BaFe$_2$As$_2$. FeSe$_{1-x}$Te$_x$ (0 $<$ $x$ $<$ 0.7) shows superconductivity
with a maximum $T_c$ $\sim$ 15 K. An extended x-ray absorption fine-structure
 (EXAFS) study has revealed that the Se and Te atoms are distributed with 
short Fe-Se bonds and long Fe-Te bonds as shown in Fig. 1(b). \cite{10} 
Since the difference in chalcogen height between FeSe and FeTe 
reverses Fe 3$d$ orbital levels as displayed in Figs. 1(c) and (d), 
the Fe-Se/Te bond disorder may result in the Fe 3$d$ orbital disorder 
in the real space. 
In this paper, we report an angle-resolved photoemission spectroscopy 
(ARPES) study on FeSe$_{1-x}$Te$_x$
which reveal the impact of ligand field reverse between FeSe and FeTe
on the fundamental electronic structure of FeSe$_{1-x}$Te$_x$.
In FeSe$_{1-x}$Te$_x$, the distribution of the Fe-Se and Fe-Te bonds in the real space
causes the splitting of the hole band with Fe 3$d$ $x^2-y^2$ character
around the $\Gamma$ point in the momentum space.
Since the hole pocket of the Fe 3$d$ $x^2-y^2$ band is expected to play 
important roles in various proposed paring mechanisms including $s_{+-}$ 
paring by spin fluctuation \cite{Kuroki2008,Mazin2008} and $s_{++}$ paring 
by orbital fluctuation, \cite{Kontani2010}
the present study on the orbital-selective band splitting due to the chalcogen height disorder 
suggests that the inhomogeneity in the orbital channel should be included 
to describe the Fe-based superconductors.

We have studied single crystals of annealed FeSe$_{1-x}$Te$_x$ with 
$x$ = 0.6 and 0.9 which were grown as reported by Noji {\it et al.} \cite{11} 
ARPES measurements were performed at beamline 9A, Hiroshima Synchrotron 
Radiation Center (HSRC) which has a normal incidence monochromator 
with off-plane Eagle mounting. The end station is equipped with 
a SCIENTA R4000 analyzer for angle-resolved photoemission experiments, 
which is a 200 mm mean radius spectrometer. 
The helical undulator and the monochromator provides circularly polarized light 
for the energy range of 4-30 eV. In the present experiment, the photon energy
was set to $h\nu$ = 17 eV and 23 eV. Total energy resolutions 
including the monochromator and the electron analyzer were set to 
18 meV and 14 meV for $h\nu$ = 23 eV and 17 eV, respectively. 
The base pressure of the spectrometer was $10^{-9}$ Pa range. 
The single crystals were properly oriented on the sample stage 
by the standard Laue measurements and were cooled using liquid He refrigerator. 
We cleaved the single crystals at 14 K under ultrahigh vacuum 
and took ARPES data at 14 K within four hours after the cleavage.

\begin{figure}
\includegraphics[width=0.35\textwidth]{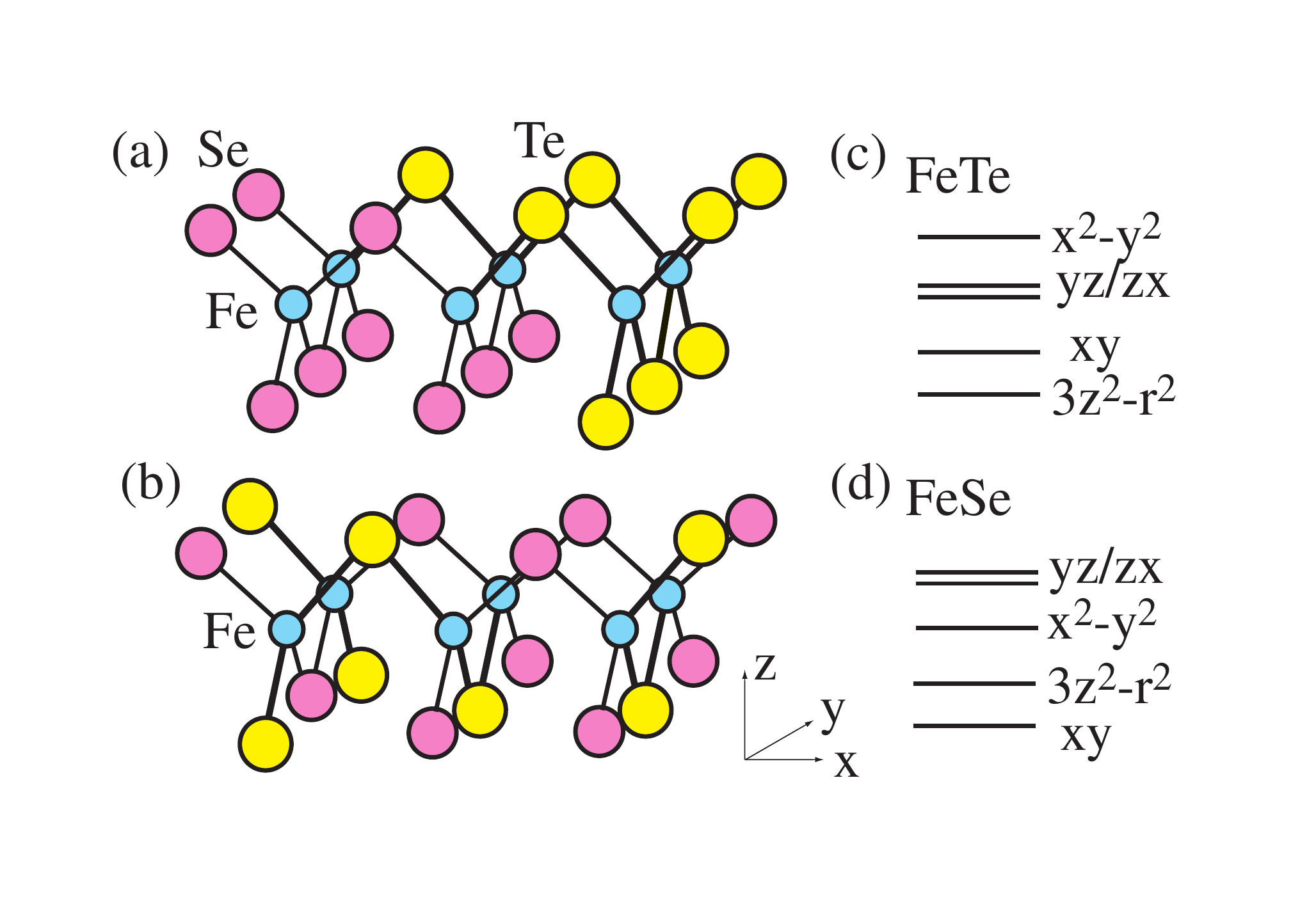}
\caption{
(Color online) (a) Crystal structure of FeSe/FeTe and
(b) that of FeSe$_{1-x}$Te$_x$ with short Fe-Se bonds and 
long Fe-Te bonds. The FeTe$_4$ tetrahedron is elongated along 
the $z$-axis compared to the FeSe$_4$ tetrahedron.
(c) Fe 3$d$ energy levels for FeTe and (d) those for FeSe.
}
\end{figure}

\begin{figure}
\includegraphics[width=0.4\textwidth]{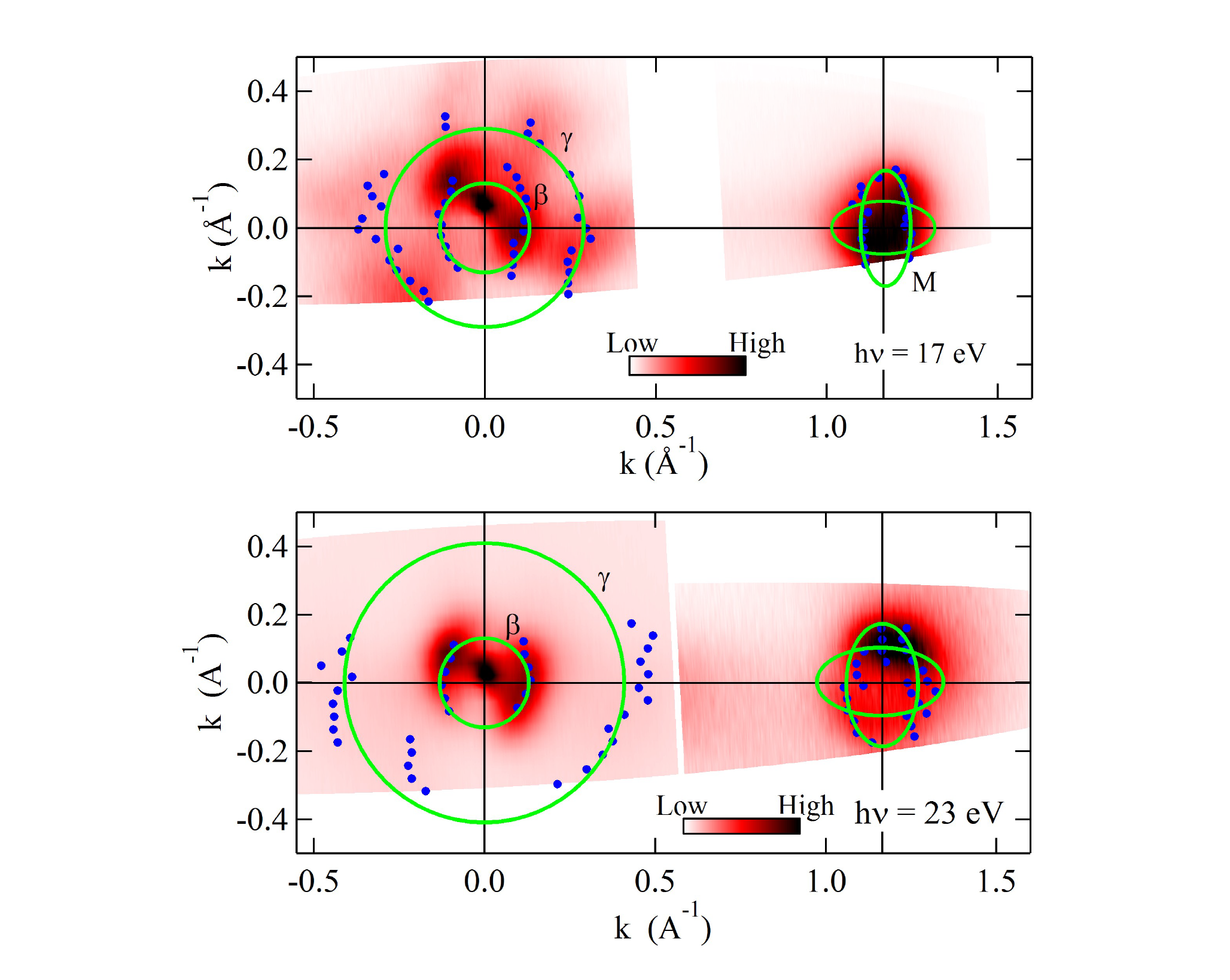}
\caption{
(Color online) Fermi surface mappings for annealed FeSe$_{0.4}$Te$_{0.6}$ 
at $h\nu$ = 17 eV and $h\nu$ = 23 eV. The dots indicate the band locations 
determined by peak positions of second derivative of MDC.
}
\end{figure}

Figure 2 shows Fermi surfaces of annealed FeSe$_{1-x}$Te$_x$ 
with $x$ = 0.6 taken at $h\nu$ = 17 eV and 23 eV. For the Fermi surface mapping, 
we used ARPES intensity integrated within an energy window of $\pm$5 meV 
at the Fermi level ($E_F$).
The momentum points where the bands cross $E_F$ can be determined by 
maximum points in second derivative spectrum of momentum distribution 
curve (MDC) at $E_F$. The momentum points thus determined are indicated 
by the dots, and the Fermi surfaces deduced from the dots are shown 
in the maps. 
The cut from $\Gamma$ (zone center) to M (zone corner) corresponds 
to the in-plane Fe-Fe direction. As commonly observed in various Fe-based
superconductors, the hole-like Fermi surfaces around $\Gamma$ and
the electron-like Fermi surface around M are observed.
The area of the outer Fermi surface increases in going from $h\nu$ = 17 eV
to $h\nu$ = 23 eV, indicating that the Fermi surface depends on the momentum
perpendicular to the FeSe plane. In the following discussion, we focus
on the ARPES data taken at $h\nu$ = 23 eV in which the $\Gamma$ point
(zone center) of the two-dimensional Brillouin zone is rather close 
the Z point of the three-dimensional Brillouin zone, and the hole bands 
are more clearly observed compared to those at $h\nu$ = 17 eV.

Figure 3(a) shows band dispersions along the $\Gamma$-M direction
of annealed FeSe$_{1-x}$Te$_x$ with $x$ = 0.6 at $h\nu$ = 23 eV. 
The band dispersions are extracted from the EDC data displayed in Fig. 3(b).
Three hole-like bands at the zone center are clearly observed. 
Among them, two outer hole-like bands cross $E_F$ and the inner hole-like band 
does not cross $E_F$ with top position at about 20 meV below $E_F$ 
judging from the energy distribution curve (EDC). (It is very difficult to 
identify the band top using the second derivative spectrum of MDC since the tail 
above the band top, which is mainly due to finite energy resolution, appears 
as a maximum point in MDC.)
This is consistent with two hole-like pockets around the zone center ($\Gamma$ point) 
observed in Fig. 2, and is qualitatively consistent with the previous ARPES 
results on FeSe$_{0.42}$Te$_{0.58}$, \cite{12} FeTe, \cite{13}
Fe$_{1.03}$Te$_{0.7}$Se$_{0.3}$, \cite{14} and FeSe$_{0.5}$Te$_{0.5}$. \cite{15} 

\begin{figure}
\includegraphics[width=0.4\textwidth]{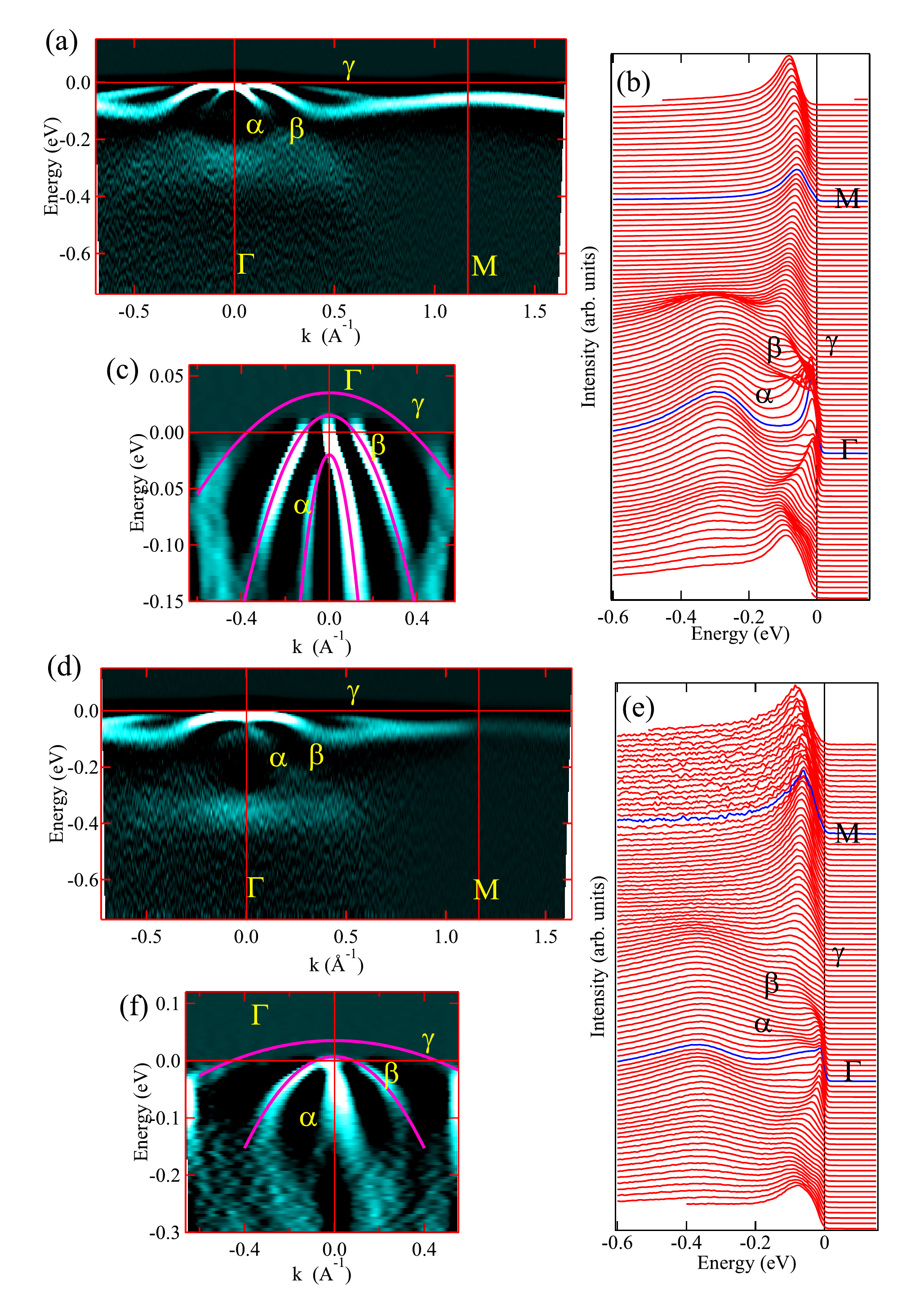}
\caption{
(Color online) ARPES results for annealed  FeSe$_{0.4}$Te$_{0.6}$
taken at $h\nu$ = 23 eV. (a) Second derivative plot of EDC. (b) EDC spectrum
along $\Gamma$-M. (c) Second derivative plot of MDC around $\Gamma$ point.
ARPES results for annealed  FeSe$_{0.1}$Te$_{0.9}$ taken at $h\nu$ = 23 eV. 
(d) Second derivative plot of EDC. (e) EDC spectrum along $\Gamma$-M. 
(f) Second derivative plot of MDC around $\Gamma$ point.
}
\end{figure}

\begin{figure}
\includegraphics[width=0.4\textwidth]{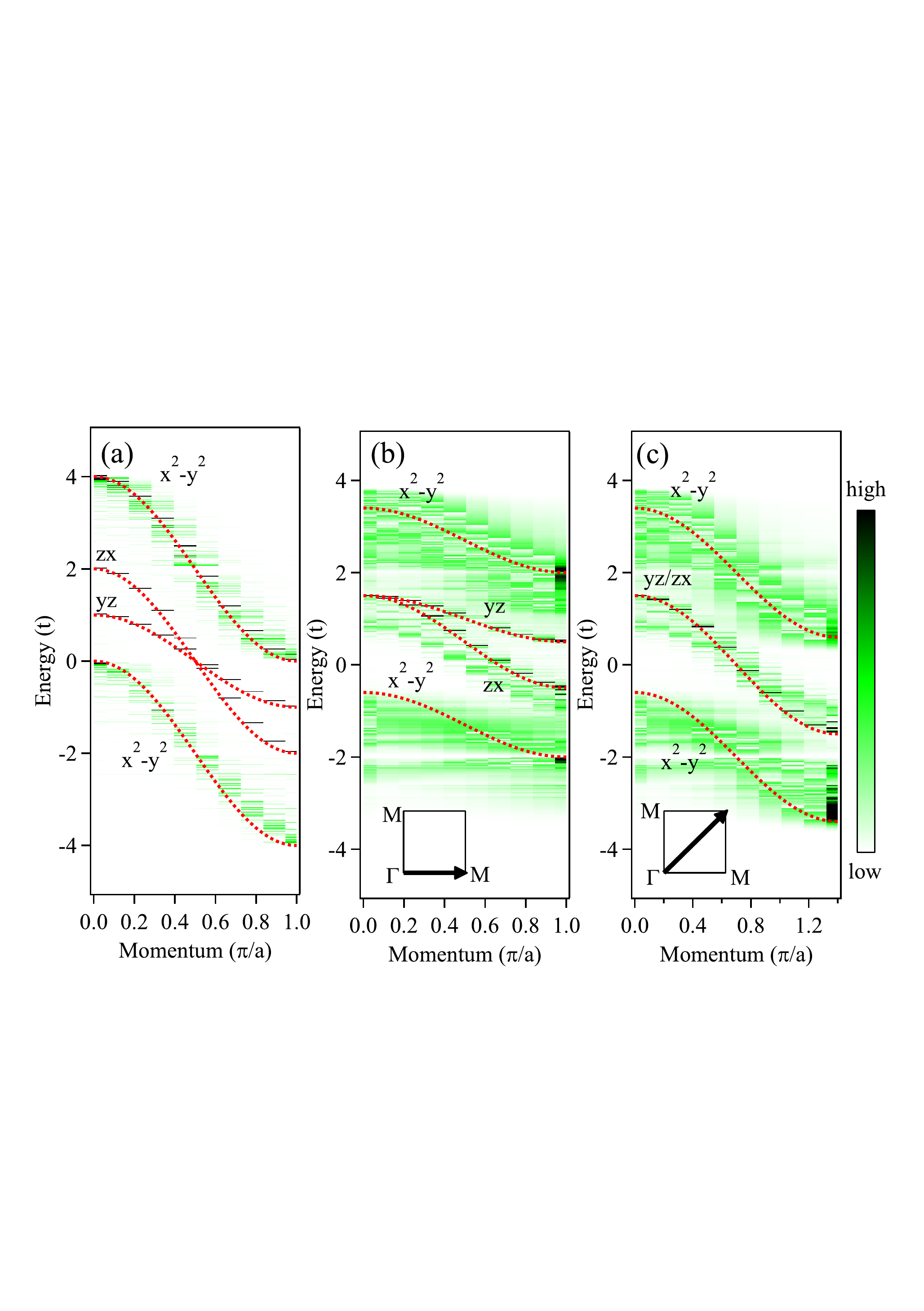}
\caption{
(Color online) Momentum distribution of single-electron excitation 
(a) for the 1D multi-band model with random orbital level splitting,
(b) for the (0,0) to ($\pi$,0) direction of 2D model, 
(c) for the (0,0) to ($\pi$,$\pi$) direction of 2D model.
Momentum and energy are given in units of $\pi/a$ and $t$ where
$a$ is the lattice constant and $t$ is the transfer integral.
}
\end{figure}

The present results show that the effect of Fe-Se/Te bond disorder
is apparently small in the momentum space. The difference of Fe-Se/Te bond, 
in particular, Se/Te height from the Fe plane is predicted to give 
substantial change of the electronic structure between FeSe and FeTe 
both in the zone center and zone corner. \cite{12} 
In the case of FeSe, there are the $x^2-y^2$ band below $E_F$ and 
only $yz/zx$ bands crossing $E_F$ around $\Gamma$ point, while, 
in the case of FeTe, there are three bands crossing $E_F$ and $x^2-y^2$
($yz/zx$) bands have higher (lower) energy around $\Gamma$. 
Although the Fe 3$d$ orbitals are completely disordered due to 
the bond disorder, the clear band structures survive around the zone center.
Here, it should be noted that the Se/Te distribution in FeSe$_{1-x}$Te$_{x}$
becomes more homogeneous or more random by annealing
as confirmed by x-ray diffraction analysis and 
electron probe micro analysis. \cite{16,17}

The momentum dependent effect of atomic disorder in random alloys 
of $AB_{1-x}C_x$, \cite{18} in which the local structure around 
the $A$ site is disordered although the end members $AB$ and $AC$ 
possessing perfect crystallographic symmetry, \cite{19} provides a hint 
to understand the ARPES results of FeSe$_{0.4}$Te$_{0.6}$.
In the strongly perturbed alloys, the band structure of $AB_{1-x}C_x$ deviates 
from the average between those of $AB$ and $AC$ due to formation of impurity bands. 
The specialty of Fe(Se,Te) is that the Fe 3$d$ level splitting is reversed 
between FeSe and FeTe. The EXAFS results \cite{10} indicate that the Fe 3$d$ 
orbital splitting locally have FeSe-like orbital state or FeTe-like orbital 
state instead of the average between them. 
In such case, the FeSe-like band dispersion and the FeTe-like band dispersion 
can coexist in the momentum space to be observed by ARPES. 

In order to support this scenario on the coexistence of FeSe-like and FeTe-like
band dispersions, momentum distribution of single-electron excitation is calculated
using a 19-site one-dimensional (1D) model and a 19 $\times$ 19-sites two-dimensional
(2D) model with random Fe 3$d$ level splitting. In the present models with 
Fe 3$d$ $x^2-y^2$, $yz$, and $zx$ orbitals, the energies of $x^2-y^2$, $yz$, 
and $zx$ are set to -2$t$ (2$t$), 0, 0 at the FeSe-like (FeTe-like) sites 
to consider the reverse of orbital splitting.
In the 1D model with lattice constant $a$, the transfer integrals between 
the neighboring $x^2-y^2$ and $zx$ orbitals are set to $t$ while those 
between the neighboring $yz$ orbitals are set to $t/2$.
In the 2D square lattice model with lattice constant $a$, 
the transfer integrals between the neighboring $xy$ orbitals are $t/2$ 
along the x- and y-axes and those between the neighboring $yz$ ($zx$) 
orbitals are $t/4$ and $t/2$ ($t/2$ and $t/4$) along the x- and y-axes respectively. 
All the configurations for $\sim$ 40\% FeSe-like sites and $\sim$ 60\% FeTe-like sites 
are included in the 1D case while 10000 configurations are randomly selected 
for the 2D case.
The calculated momentum distribution for the 1D model is displayed in Fig. 4(a)
where the Fe 3$d$ $x^2-y^2$ band is split into the two bands which correspond to 
the FeTe-like and FeSe-like band dispersions (roughly follow $\epsilon_k = 
\pm2t + 2t\cos(k_x)$ as indicated by the dotted curves respectively).
The splitting of the Fe 3$d$ band is also reproduced in the 2D model 
along the (0,0) to ($\pi/a$,0) direction [Fig. 4( b)] and 
along the (0,0) to ($\pi/a$,$\pi/a$) direction [Fig. 4(c)]. 
In the 2D case, the band dispersions of the Fe 3$d$ $x^2-y^2$ bands are roughly
fit to $\epsilon_k = \pm2t + 0.7t[\cos(k_x)+\cos(k_y)]$ (shown by the dotted curves) 
indicating that renormalization effect with factor of 0.7. The renormalization effect
due to the random orbital distribution could contribute to the enhancement
of electronic specific heat in FeSe$_x$Te$_{1-x}$.

The ARPES results of FeSe$_{0.4}$Te$_{0.6}$ can be interpreted 
on the basis of the present calculations.
The inner hole band labeled as $\alpha$ can be assigned to 
the $x^2-y^2$ band of the FeSe-like orbital state. 
On the other hand, the outer hole band labeled as $\gamma$ is derived 
from the $x^2-y^2$ band of the FeTe-like orbital state. 
The $yz/zx$ bands of FeSe- and FeTe-like orbital states, which are 
almost degenerate in the LDA calculations, \cite{12} can be merged 
as demonstrated in the above calculations and can be observed as $\beta$.
At $x$ = 0.6, the annealing induces small increase of $T_c$ \cite{11} which is 
probably due to strain relaxing atomic redistribution for the two kinds of 
orbital states, suppressing the magnetic order of FeTe-like orbital domains
(compressive for the one and tensile for the other domains) as happens 
in various granular superconductors and random alloys. \cite{18} 
At this point, it is very interesting and important to study $x$ = 0.9 
in which the as-grown sample is not superconducting and the annealing 
induces $T_c$ $\sim$ 11 K. \cite{11}

Figure 3(d) shows band dispersions of annealed FeSe$_{1-x}$Te$_x$ 
with $x$ = 0.9 at $h\nu$ = 23 eV. The band dispersions are
extracted from the EDC data displayed in Figs. 3(e).
At least two hole-like bands at the zone center are observed. 
The Te rich sample is dominated by the FeTe-like orbital states.
Therefore, it is expected that the $x^2-y^2$ band (labeled as $\gamma$) 
is higher in energy than the $yz/zx$ band at the zone center 
and the $x^2-y^2$ band is smoothly connected to the flat band 
at the zone corner.
This expectation is consistent with the observed band dispersions
shown in Figs. 3(d) and (f). 
In addition, the width of the outer hole band
labeled as $\gamma$ is smaller at $x$ = 0.9 than $x$ = 0.6.
Such band structure of FeTe is known to support magnetic state 
instead of superconductivity. 
However, at $x$ = 0.9, the Se/Te distribution helps to form FeSe-like orbital site 
where the $yz/zx$ band (labeled as $\beta$) is higher in energy 
than the $x^2-y^2$ band (labeled as $\alpha$). 
Here, it should be noted that the intensity of band $\alpha$ is very small 
at $x$ = 0.9 compared to that at $x$ = 0.6, consistent with the assumption
that it is derived from the minor FeSe-like orbital state. 
At this stage. one can speculate that the superconductivity appears 
at $x$ = 0.9 due to strain relaxing distribution of the domains with different 
orbital states on annealing.

In conclusion, we have reported an ARPES study on the annealed 
FeSe$_{1-x}$Te$_x$ ($x$ = 0.6 and 0.9). 
The comparison between the band dispersions at $x$ = 0.6 
and those at $x$ = 0.9 shows that the disorder of Fe 3$d$ level splitting 
affects the Fe 3$d$ band dispersions in an orbital-selective way.
The hole bands at the zone center in FeSe$_{1-x}$Te$_x$ can be assigned 
to the Fe 3$d$ bands from the FeSe-like and the FeTe-like orbital states
as supported by the model calculation. At $x$ = 0.9, the minor FeSe-like 
orbital state may induce the superconductivity due to strain relaxing 
atomic distribution under the atomic disorder.

The synthesis and characterization of the FeSe$_{1-x}$Te$_x$ single crystals were 
supported by a Grant-in-Aid for Scientific Research from the Japan Society 
for Promotion of Science. The synchrotron radiation experiments have been done 
with the approval of HSRC (Proposal No. 10-A-10).


\begin{thebibliography}{99}

\bibitem{1}
J. M. Tranquada, B. J. Sternlieb, J. D. Axe, Y. Nakamura, S. Uchida, 
Nature {\bf 375}, 561 (1995).

\bibitem{2}
A. Bianconi, N. L. Saini, A. Lanzara, M. Missori,  T. Rossetti, H. Oyanagi, H. Yamaguchi, K. Oka, T. Ito, 
Phys. Rev. Lett. {\bf 76}, 3412 (1996).

\bibitem{3}
K. M. Lang, V. Madhavan, J. E. Hoffman, E. W. Hudson, H. Eisaki, S. Uchida,
Nature {\bf 415}, 412 (2002).

\bibitem{7}
Y. Kamihara, T. Watanabe, M. Hirano, and H. Hosono,
J. Am. Chem. Soc. {\bf 130}, 3296 (2008). 

\bibitem{8}
F. C. Hsu, J. Y. Luo, K. W. The, T. K. Chen, T. W. Huang, P. M. Wu, Y. C., Lee, Y. L. Huang, 
Y. Y. Chu, D. C. Yan, M. K. Wu, Proc. Natl. Acad. Sci. U.S.A. {\bf 105}, 14262 (2008).

\bibitem{9}
W. Bao, Y. Qiu, Q. Huang, M. A. Green, P. Zajdel, M. R. Fitzsimmons, M. Zhernenkov, S. Chang, M. Fang, 
B. Qian, E. K. Vehstedt, J. Yang, H. M. Pham, L. Spinu, Z. Q. Mao,
Phys. Rev. Lett. {\bf 102}, 247001 (2009).

\bibitem{10}
B. Joseph, A. Iadecola, A. Puri, L. Simonelli, Y. Mizuguchi, Y. Takano, N. L. Saini, 
Phys. Rev. B {\bf 82}, 020502 (2010).

\bibitem{Kuroki2008}
K. Kuroki, S. Onari, R. Arita, H. Usui, Y. Tanaka, H. Kontani, and H. Aoki,
Phys. Rev. Lett. {\bf 101}, 087004 (2008).

\bibitem{Mazin2008}
I. I. Mazin, D. J. Singh, M. D. Johannes, and M. H. Du, 
Phys. Rev. Lett. {\bf 101}, 057003 (2008).

\bibitem{Kontani2010}
H. Kontani and S. Onari,
Phys. Rev. Lett. {\bf 104}, 157001 (2010).

\bibitem{11}
T. Noji, T. Suzuki, H. Abe, T. Adachi, M. Kato, Y. Koike, 
J. Phys.Soc. Jpn. {\bf 79}, 084711 (2010).

\bibitem{12}
A. Tamai, A. Y. Ganin, E. Rozbicki, J. Bacsa, W. Meevasana, P. D. C. King, 
M. Caffio, R. Schaub, S. Margadonna, K. Prassides, M. J. Rosseinsky, F. Baumberger, 
Phys. Rev. Lett. {\bf 104}, 097002 (2010).

\bibitem{13}
Y. Xia, D. Qian, L. Wray, D. Hsieh, G. F. Chen, J. L. Luo, N. L. Wang, M. Z. Hasan, 
Phys. Rev. Lett. {\bf 103}, 037002 (2009).

\bibitem{14}
K. Nakayama, T. Sato, P. Richard, T. Kawahara, Y. Sekiba, T. Qian, G. F. Chen, J. L. Luo, 
N. L. Wang, H. Ding, T. Takahashi,
Phys. Rev. Lett. {\bf 105}, 197001 (2010).

\bibitem{15}
S.-H. Lee, G. Xu, W. Ku, J. S. Wen, C. C. Lee, N. Katayama, Z. J. Xu, S. Ji, Z. W. Lin, G. D. Gu,
H.-B. Yang, P. D. Johnson, Z.-H. Pan, T. Valla, M. Fujita, T. J. Sato, S. Chang, K. Yamada, 
J. M. Tranquada, Phys. Rev. B {\bf 81}, 220502(R) (2010).

\bibitem{16}
M. Imaizumi, T. Noji, T. Adachi, Y. Koike, Physica C {\bf 471}, 614 (2011).

\bibitem{17}
T. Taen, Y. Tsuchiya, Y. Nakajima, T. Tamegai, 
Phys. Rev. B {\bf 80}, 092502 (2009).

\bibitem{18}
V. Popescu and A. Zunger, Phys. Rev. Lett. {\bf 104}, 236403 (2010).

\bibitem{19}
P. F. Peterson, Th. Proffen, I.-K. Jeong, S. J. L. Billinge, K.-S. Choi, M. G. Kanatzidis, P. G. Radaelli,
Phys. Rev. B {\bf 63}, 165211 (2001).

\end{thebibliography}
\end{document}